\documentclass[sigconf]{acmart}

\usepackage{textcomp}
\usepackage{stfloats}
\usepackage{url}
\usepackage{verbatim}
\usepackage{amsthm}
\usepackage{subfigure}
\usepackage{makecell}
\usepackage{multirow}
\usepackage{enumitem}
\usepackage{subcaption}

\usepackage{color, colortbl}

\definecolor{Gray}{rgb}{0.9, 0.9, 0.9}

\usepackage[capitalize,noabbrev]{cleveref}

\newtheorem{theorem}{Theorem}
\newtheorem{lemma}{Lemma}

\newtheorem{corollary}{Corollary}

\usepackage[ruled,linesnumbered]{algorithm2e}

\usepackage[inkscapelatex=false]{svg}

\AtBeginDocument{%
  }

\copyrightyear{2025}
\acmYear{2025}
\setcopyright{acmlicensed}\acmConference[RecSys '25]{Proceedings of the Nineteenth ACM Conference on Recommender Systems}{September 22--26, 2025}{Prague, Czech Republic}
\acmBooktitle{Proceedings of the Nineteenth ACM Conference on Recommender Systems (RecSys '25), September 22--26, 2025, Prague, Czech Republic}
\acmDOI{10.1145/3705328.3748059}
\acmISBN{979-8-4007-1364-4/2025/09}

\begin{document}


\title[NLGCL]{NLGCL: Naturally Existing Neighbor Layers Graph Contrastive Learning for Recommendation}

\author{Jinfeng Xu}
\email{jinfeng@connect.hku.hk}
\affiliation{%
  \institution{The University of Hong Kong}
  \city{HongKong SAR}
  \country{China}}

\author{Zheyu Chen}
\email{zheyu.chen@connect.polyu.hk}
\affiliation{%
  \institution{The Hong Kong Polytechnic University}
  \city{HongKong SAR}
  \country{China}}

\author{Shuo Yang}
\email{shuoyang.ee@gmail.com}
\affiliation{%
  \institution{The University of Hong Kong}
  \city{HongKong SAR}
  \country{China}}

\author{Jinze Li}
\email{lijinze-hku@connect.hku.hk}
\affiliation{%
  \institution{The University of Hong Kong}
  \city{HongKong SAR}
  \country{China}}

\author{Hewei Wang}
\email{heweiw@andrew.cmu.edu}
\affiliation{%
    \institution{Carnegie Mellon University}  
    \city{Pittsburgh, PA}   
    \country{USA}}

\author{Wei Wang}
\email{ehomewang@ieee.org}
\affiliation{%
  {\institution{Shenzhen MSU-BIT University}}
  \city{Shenzhen}
  \country{China}}

\author{Xiping Hu}
\email{huxp@bit.edu.cn}
\affiliation{%
  {\institution{Beijing Institute of Technology}}
  \city{Beijing}
  \country{China}}

\author{Edith Ngai}
\authornote{Corresponding authors}
\email{chngai@eee.hku.hk}
\affiliation{%
  \institution{The University of Hong Kong}
  \city{HongKong SAR}
  \country{China}}

\renewcommand{\shortauthors}{Xu et al.}

\begin{abstract}
Graph Neural Networks (GNNs) are widely used in collaborative filtering to capture high-order user-item relationships. To address the data sparsity problem in recommendation systems, Graph Contrastive Learning (GCL) has emerged as a promising paradigm that maximizes mutual information between contrastive views. However, existing GCL methods rely on augmentation techniques that introduce semantically irrelevant noise and incur significant computational and storage costs, limiting effectiveness and efficiency.

To overcome these challenges, we propose NLGCL, a novel contrastive learning framework that leverages naturally contrastive views between neighbor layers within GNNs. By treating each node and its neighbors in the next layer as positive pairs, and other nodes as negatives, NLGCL avoids augmentation-based noise while preserving semantic relevance. This paradigm eliminates costly view construction and storage, making it computationally efficient and practical for real-world scenarios. Extensive experiments on four public datasets demonstrate that NLGCL outperforms state-of-the-art baselines in effectiveness and efficiency.
\end{abstract}



\maketitle

\section{Introduction}
The proliferation of the internet has led to an explosion of information, making recommender systems indispensable in modern society, including domains such as social media \cite{godin2013using,chen2025squeeze,xu2025survey}, e-commerce \cite{smith2017two,chen2025don,xu2025cohesion}, and short video platforms \cite{covington2016deep,xu2025mdvt}. Traditional recommender systems typically model user preferences based on historical user-item interactions \cite{he2020lightgcn,xu2024improving}. With advances in graph representation learning \cite{kipf2017semi,wu2019simplifying}, Graph Neural Networks (GNNs) have been introduced to extract collaborative signals from high-order neighbors, enhancing representation learning in recommender systems \cite{he2020lightgcn,liu2021interest,zhou2023layer,wang2019neural}. Despite their popularity and effectiveness, learning high-quality user and item representations remains challenging due to the data sparsity problem and label dependency. To mitigate data sparsity, self-supervised learning (SSL) offers a promising avenue by generating supervised signals from unlabeled data \cite{liu2021self,liu2022graph}. Among SSL paradigms, contrastive learning (CL) \cite{jing2023contrastive} has shown potential by maximizing mutual information between positive pairs in contrastive views, thereby leveraging self-supervised signals to enhance learning. Recently, graph contrastive learning (GCL) methods \cite{wu2021self,yu2023xsimgcl,yu2022graph} have garnered significant attention in the recommender system field. Typically, GCL methods construct contrastive views using data augmentation techniques such as node deletion, edge perturbation, feature masking, and noise addition \cite{rebuffi2021data,bayer2022survey}. For example, SGL \cite{wu2021self} applies stochastic perturbations to nodes and edges, while SimGCL \cite{yu2022graph} and LightGCL \cite{cailightgcl} manipulate graph structures via noise addition and singular value decomposition, respectively. Other methods like NCL \cite{lin2022improving} use clustering algorithms, and DCCF \cite{ren2023disentangled} and BIGCF \cite{zhang2024exploring} focus on disentangled representations through SSL.

However, existing GCL methods have significant limitations regarding both \textbf{effectiveness} and \textbf{efficiency}, which are discussed and validated in Section~\ref{sec: investigation} by comprehensive investigation. \textbf{Effectiveness}: While augmentation techniques enable models to learn robust and invariant features, they can introduce semantically irrelevant noise \cite{wu2021self,yu2022graph} that distorts critical structural and feature information. Random alterations may not preserve the underlying semantics of the graph, potentially hindering model performance \cite{lin2022improving,ren2023disentangled}. \textbf{Efficiency}: Constructing and storing augmented contrastive views increases computational and storage overhead, leading to higher time complexity and reduced scalability \cite{cailightgcl,zhang2024exploring}. In practical scenarios where quick responses and resource efficiency are crucial, these additional costs limit the applicability of existing methods.

    

To this end, we propose a simple yet effective paradigm called Neighbor Layers Graph Contrastive Learning (NLGCL). Unlike existing GCL methods relying on data augmentation, NLGCL leverages naturally existing contrastive views within GNNs. Specifically, NLGCL treats each node and its neighbors in the next layer as positive pairs, and each node and other nodes as negative pairs. In the recommendation field, graphs are constructed from historical user-item interactions. Users and items with similar interaction histories share similar semantic information after neighbor aggregation in GNNs. By utilizing the naturally existing contrastive views between neighbor layers, NLGCL avoids introducing irrelevant noise and eliminates the need for additional augmented views, reducing computational costs. We conduct extensive experiments to validate the superiority of our NLGCL over various state-of-the-art baselines in terms of effectiveness and efficiency. 
\section{Preliminary}
\subsection{Definition}
In this section, we provide some essential notations and definitions. The historical user-item interactions can be naturally represented as a bipartite graph $\mathcal{G}$ $=$ $(\mathcal{V}, \mathcal{E})$. The node set $\mathcal{V}$ includes user nodes $\mathcal{U} = \{u_1, u_2, ..., u_{|\mathcal{U}|}\}$ and item nodes $\mathcal{I} = \{i_1, i_2, ..., i_{|\mathcal{I}|}\}$. The edge set $\mathcal{E}$ consists of undirected edges between user nodes and item nodes. The user-item interaction matrix is denoted as $\mathcal{R} \in \{0,1\}^{|\mathcal{U}| \times |\mathcal{I}|}$. Specifically, each entry $\mathcal{R}_{u,i}$ indicates whether the user $u$ is connected to item $i$, with a value of 1 representing a connection and 0 otherwise. The entire user-item interaction matrix can be divided into an observed user-item interaction matrix $\mathcal{R^{+}} \in \{\mathcal{R}_{u,i}|u \in \mathcal{U},i \in \mathcal{I},\mathcal{R}_{u,i} = 1\}$ and an unobserved user-item interaction matrix $\mathcal{R^{-}} \in \{\mathcal{R}_{u,i}|u \in \mathcal{U},i \in \mathcal{I},\mathcal{R}_{u,i} = 0\}$. It is clear that the number of undirected edges $|\mathcal{E}|$ equals the number of observed user-item interactions $|\mathcal{R^{+}}|$ in the training data. We random initialize $\mathbf{E}_u \in \mathbb{R}^{d \times |\mathcal{U}|}$ and $\mathbf{E}_i \in \mathbb{R}^{d \times |\mathcal{I}|}$ to represent user and item embeddings, respectively. The graph structure of $\mathcal{G}$ can be denoted as the adjacency matrix $\mathcal{A} \in \mathbb{R}^{(|\mathcal{U}|+|\mathcal{I}|) \times (|\mathcal{U}|+|\mathcal{I}|)}$, formally: 
\vskip -0.15in
\begin{equation}
\mathcal{A}=\left[\begin{array}{cc}
0^{|\mathcal{U}| \times |\mathcal{U}|} & \mathcal{R} \\
 \mathcal{R}^T & 0^{|\mathcal{I}| \times |\mathcal{I}|}
\end{array}\right].
\end{equation}
\vskip -0.1in

The symmetrically normalized matrix is $\mathcal{\tilde{A}}$ $=$ $\mathcal{D}^{-\frac{1}{2}}\mathcal{A}\mathcal{D}^{-\frac{1}{2}}$, where $\mathcal{D}$ represents diagonal degree matrix.




\subsection{GNNs for Recommendation}
GNNs update node representation through aggregating messages from their neighbors. The core of the graph-based CF paradigm consists of two steps: \textbf{S1: message propagation} and \textbf{S2: node representation aggregation}. Thus, message propagation for the user/item node can be formulated as: $\mathbf{e}_u^{(l)}=\operatorname{Aggr}^{(l)}(\{\mathbf{e}_i^{(l-1)}: i \in \mathcal{N}_u\})$ and $\mathbf{e}_i^{(l)}=\operatorname{Aggr}^{(l)}(\{\mathbf{e}_u^{(l-1)}: u \in \mathcal{N}_i\})$, where $\mathcal{N}_u$ and $\mathcal{N}_i$ denote the neighborhood set of nodes $u$ and $i$, respectively, and $l$ denotes the $l$-th layer of GNNs. Then, the final user/item embeddings can be formulated as: $\mathbf{\bar{E}}_u=\operatorname{Readout}([\mathbf{E}_u^{(0)},\mathbf{E}_u^{(1)},...,\mathbf{E}_u^{(L)}])$ and $\mathbf{\bar{E}}_i=\operatorname{Readout}([\mathbf{E}_i^{(0)},\mathbf{E}_i^{(1)},...,\mathbf{E}_i^{(L)}])$, where the $\operatorname{Readout}(\cdot)$ function can be any differentiable function, and $L$ is the layer number of GNN. In the recommendation scenario, LightGCN \cite{he2020lightgcn} is currently the most popular GNN backbone, which effectively captures high-order information through neighborhood aggregation. For efficient training and to ensure that diverse neighborhood information is integrated, final embeddings are aggregated from all layers: $\mathbf{\bar{E}}_u=\frac{1}{L+1}(\mathbf{E}_u^{(0)}+\mathcal{\tilde{A}}^1 \mathbf{E}_u^{(0)}+...+\mathcal{\tilde{A}}^L \mathbf{E}_u^{(0)})$ and $
\mathbf{\bar{E}}_i=\frac{1}{L+1}(\mathbf{E}_i^{(0)}+\mathcal{\tilde{A}}^1 \mathbf{E}_i^{(0)}+...+\mathcal{\tilde{A}}^L \mathbf{E}_i^{(0)})$, where $\mathbf{\bar{E}}_u$ and $\mathbf{\bar{E}}_i$ denote final embeddings for user and item, respectively. Subsequently, we derive scores for all unobserved user-item pairs using the inner product of the final embeddings of the user and item, denoted as $y_{u,i}=\mathbf{\bar{e}}_u^{\top} \mathbf{\bar{e}}_i$, where $\mathbf{\bar{e}}_u$ and $\mathbf{\bar{e}}_i$ represent the final representations of user $u$ and item $i$, respectively. The items with the top-$N$ highest score are recommended to the user.

To provide effective item recommendations from user-item interactions, a typical training objective is the pair-wise loss function. We take the most widely adopted BPR \cite{rendle2012bpr} loss as an example:
\vskip -0.1in
\begin{equation}
\label{eq: bpr}
\mathcal{L}_{bpr}=\sum_{(u,p,n) \in \mathcal{O}}-\ln \sigma(y_{u,p}-y_{u,n}),
\end{equation}
\vskip -0.05in
\noindent where $\mathcal{O}=\{(u,p,n) \mid(u,p) \in \mathcal{R}^{+},(u,n) \in \mathcal{R}^{-}\}$ denotes the pair-wise training data, $\sigma(\cdot)$ denotes sigmoid function. Essentially, BPR aims to widen the predicted preference margin between the positive item $p$ and negative item $n$ for user $u$.
\subsection{CL for Recommendation}
Recent studies \cite{yu2023xsimgcl,yu2022graph,wu2021self,cailightgcl} have demonstrated that CL, through the generation of self-supervised signals, effectively mitigates the challenge of the data sparsity problem in recommender systems. CL-based methods construct contrastive views through various data augmentation strategies and optimize the mutual information between contrastive views, thereby obtaining self-supervised signals. By maximizing the consistency among positive samples and minimizing the consistency among negative samples, most of the existing CL methods mainly adopt InfoNCE \cite{oord2018representation} for optimization:
\vskip -0.15in
\begin{equation}
\label{eq: cl}
\mathcal{L}_{cl} = -\sum_{u \in \mathcal{U}}\log \frac{\exp({\frac{\mathbf{\hat{e}}_{u}^{\top} \mathbf{\tilde{e}}_{u}}{\tau}})}{\sum\limits_{v \in \mathcal{U}} \exp({\frac{\mathbf{\hat{e}}_{u}^{\top} \mathbf{\tilde{e}}_{v}}{\tau}})} - \sum_{i \in \mathcal{I}}\log \frac{\exp({\frac{\mathbf{\hat{e}}_{i}^{\top} \mathbf{\tilde{e}}_{i}}{\tau}})}{\sum\limits_{j \in \mathcal{I}} \exp({\frac{\mathbf{\hat{e}}_{i}^{\top} \mathbf{\tilde{e}}_{j}}{\tau}})},
\end{equation}
\vskip -0.05in
\noindent where $\mathbf{\hat{e}}_{u}$/$\mathbf{\hat{e}}_{i}$ and $\mathbf{\tilde{e}}_{u}$/$\mathbf{\tilde{e}}_{i}$ represent representation of item/user in different contrastive views. $\tau$ denotes the temperature hyper-parameter. However, existing CL-based approaches inevitably require the construction of multiple views, which imposes significant computational overheads. Therefore, we raise a question: \textbf{Is it necessary to construct contrastive views?} To answer this question, we provide an investigation in Section~\ref{sec: investigation}.

\section{Investigation}
\label{sec: investigation}
In this section, we analyze existing CL methods and their computational costs, identify key limitations, and emphasize the naturally existing contrastive views in GNNs. These views eliminate the need for costly view construction while preserving essential information and avoiding irrelevant noise.

\subsection{Computational Cost Analysis}
\label{sec: costs analysis}
Existing studies on CL emphasize the critical role of data augmentation in creating diverse contrastive views necessary for effective implementation \cite{wu2021self,lin2022improving,yu2022graph,yu2023xsimgcl,cailightgcl,zhang2024exploring,zhang2024recdcl}. In recommendation, there are two types: (1) graph augmentation and (2) representation augmentation. Graph augmentation typically involves generating augmented graphs by randomly discarding edges or nodes from the original graph: $\mathcal{G}'$ $=$ $\operatorname{Dropout}(\mathcal{G}(\mathcal{V},\mathcal{E}), p)$, where $\mathcal{G}$ denotes the original graph and $\mathcal{G}'$ is the augmented graph, and $p$ represents the keep rate. Representation augmented introduces noise into the graph convolution process: $\mathbf{e}'$ $=$ $\mathbf{e} + \Delta$, where $\mathbf{e}$ denotes embedding, $\Delta$ denotes the added noise. While these contrastive views are effective, data augmentation inevitably compromises the quality of embeddings: (1) Graph augmentation methods, which discard edges when generating contrastive views, result in relatively poor embedding features and density distributions. (2) Representation augmentation methods introduce noise into graph convolutions. Although the impact of some noises on distributions is less affected than that of graph augmentation, it still leads to less smooth density distributions. Furthermore, data augmentation is computationally intensive: (1) Graph augmentation methods require generating multiple augmented graphs, which is highly time-consuming. (2) Representation augmentation methods necessitate repeated graph convolution or modeling operations to obtain contrastive views, further increasing computational time. In Table~\ref{tab: efficiency}, we empirically analyze the time consumption of advanced CL-based models, supporting our observations.

\subsection{Contrastive Learning Analysis}
\label{sec: 3B}
Constructing contrastive views is the primary source of computational cost in CL-based models. Therefore, we analyze the role of view construction in contrastive learning. Existing works \cite{yu2023xsimgcl,yu2022graph} suggest that CL enables the recommendation to learn more uniformly distributed user and item representations, thereby mitigating the prevalent popularity bias in CF. Specifically, the added noise in the contrastive views propagates as part of the gradient. The uniformity is then regularized to a higher level by sampling the noise from a uniform distribution. However, we state that existing CL methods commonly face two serious problems. 

\noindent\textbf{P1}: As shown in Figure~\ref{fig: overview} (a), existing CL methods on two contrastive views can be seen as pulling each node closer to its representation in the other view and arbitrarily pushing it farther away from all other nodes in the other view, which can lead to nodes that have similar semantics being irrationally pushed farther away.

\noindent\textbf{P2}: Existing CL methods construct contrastive views by adding random noise, which inevitably discards crucial information and introduces irrelevant noise during view construction.

To address these challenges, we aim to compensate for the representation of similar nodes (\textbf{P1}) and construct semantically relevant contrastive views (\textbf{P2}). Interestingly, we find that multiple groups of such contrastive views naturally exist within GCNs. These views not only share explicit semantic relevance but also implicitly compensate for the representation of similar nodes during training.

\subsection{Naturally Contrastive Views within GCN}
\label{sec: ncv}
We point out that heterogeneous nodes in neighbor layers constitute contrastive views (e.g., ($\mathbf{E}_u^{(0)}$; $\mathbf{E}_i^{(1)}$) and ($\mathbf{E}_i^{(0)}$; $\mathbf{E}_u^{(1)}$)). To substantiate this, we first formulate the message-passing in GCNs:
\vskip -0.15in
\begin{equation}
\label{eq: neighbor1}
\mathbf{e}_u^{(l)}=\sum_{\tilde{i} \in \mathcal{N}_u} \frac{\mathbf{e}_{\tilde{i}}^{(l-1)}}{\sqrt{\left|\mathcal{N}_u\right|\left|\mathcal{N}_{\tilde{i}}\right|}}, \quad \mathbf{e}_i^{(l)}=\sum_{\tilde{u} \in \mathcal{N}_i} \frac{\mathbf{e}_{\tilde{u}}^{(l-1)}}{\sqrt{\left|\mathcal{N}_i\right|\left|\mathcal{N}_{\tilde{u}}\right|}},
\end{equation}
\vskip -0.05in
\noindent where $\mathcal{N}_u$ and $\mathcal{N}_i$ denote neighbor sets for user $u$ and item $i$, respectively. $\mathcal{N}_{\tilde{i}}$ and $\mathcal{N}_{\tilde{u}}$ denote neighbor sets for item $\tilde{i} \in \mathcal{N}_u$ and user $\tilde{u} \in \mathcal{N}_i$, respectively. Note that all users in $\mathcal{N}_{\tilde{i}}$ have the same interacted item $\tilde{i}$ with user $u$ and all items in $\mathcal{N}_{\tilde{u}}$ have the same interacted user $\tilde{u}$ with item $i$. Then we rewrite the message-passing in GCNs:
\vskip -0.18in
\begin{equation}
\label{eq: neighbor2}
\mathbf{e}_{\tilde{i}}^{(l)}=\sum_{\bar{u} \in \mathcal{N}_{\tilde{i}}} \frac{\mathbf{e}_{\bar{u}}^{(l-1)}}{\sqrt{\left|\mathcal{N}_{\tilde{i}}\right|\left|\mathcal{N}_{\bar{u}}\right|}}, \quad \mathbf{e}_{\tilde{u}}^{(l)}=\sum_{\bar{i} \in \mathcal{N}_{\tilde{u}}} \frac{\mathbf{e}_{\bar{i}}^{(l-1)}}{\sqrt{\left|\mathcal{N}_{\tilde{u}}\right|\left|\mathcal{N}_{\bar{i}}\right|}}.
\end{equation}
\vskip -0.05in

\begin{figure*}
    \centering
    \includegraphics[width=1\linewidth]{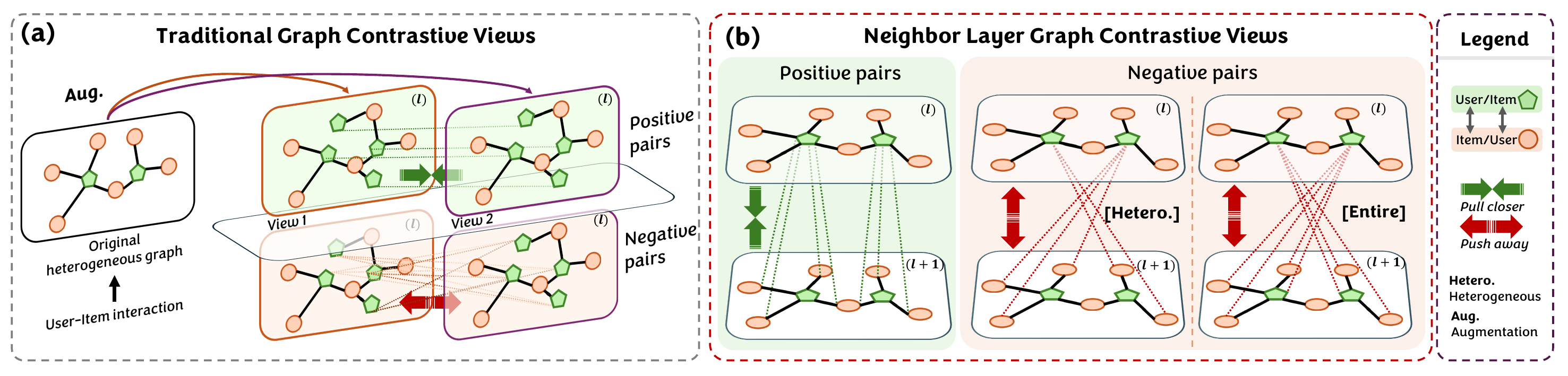}
    \caption{Overview of contrastive learning. Left: traditional contrastive learning paradigm; Right: our NLGCL.}
    \label{fig: overview}
\end{figure*}

Since item $\tilde{i} \in \mathcal{N}_u$ and user $\tilde{u} \in \mathcal{N}_i$, user $u$ and $i$ are involved in $\mathcal{N}_{\tilde{i}}$ and $\mathcal{N}_{\tilde{u}}$, respectively. By associating Eq~\eqref{eq: neighbor1} and Eq~\eqref{eq: neighbor2}, we know that representation $\mathbf{e}_{\tilde{i}}^{(l)}$ of item $\tilde{i}$ in $(l)$-th layer can be regarded as a weighted aggregation of the representation $\mathbf{e}_u^{(l-1)}$ of user $u$ in $(l$$-$$1)$-th layer and the representation $\mathbf{e}_{\bar{u}}^{(l-1)}$ of other users $\bar{u} \in \mathcal{N}_{\tilde{i}}$ in $(l$$-$$1)$-th layer who interact with item $\tilde{i}$. In contrast, representation $\mathbf{e}_{\tilde{u}}^{(l)}$ of user $\tilde{u}$ in $(l)$-th layer can be regarded as an equal-weighted aggregation of the representation $\mathbf{e}_i^{(l-1)}$ of item $i$ in $(l$$-$$1)$-th layer and the representation $\mathbf{e}_{\bar{i}}^{(l-1)}$ of other items $\bar{i} \in \mathcal{N}_{\tilde{u}}$ in $(l$$-$$1)$-th layer who interact with user $\tilde{u}$. Users with similar preferences generally share a part of semantics, aligning with the GCN's goal to enhance representations by aggregating neighbor nodes. Thus the representations $\mathbf{e}_u^{(l-1)}$ and $\mathbf{e}_i^{(l-1)}$ of user $u$ and item $i$ in $(l$$-$$1)$-th layer form positive pairs with the representations $\mathbf{e}_{\tilde{i}}^{(l)}$ and $\mathbf{e}_{\tilde{u}}^{(l)}$ of their neighbor nodes items $\tilde{i} \in \mathcal{N}_u$ and users $\tilde{u} \in \mathcal{N}_i$ in $(l)$-th layer, respectively. Thus, we find that heterogeneous nodes in neighbor layers naturally constitute contrastive views. For each $(l$$-$$1)$-th layer user $u$, the items $\tilde{i} \in \mathcal{N}_u$ in $(l)$-th layer that interact with it are treated as positive samples, and the other items $\hat{i} \in (\mathcal{I} \setminus \mathcal{N}_u )$ in $(l)$-th layer that without interact with it are treated as negative samples. Similar tendencies are presented for items. For a clearer understanding, we present the naturally existing contrastive views between neighbor layers in Figure~\ref{fig: overview} (b).

We argue that leveraging natural contrastive views removes computational overhead and addresses the limitations of existing CL methods in Section~\ref{sec: 3B}. For \textbf{P1}, in the $(l$$-$$1)$-th layer view, positive samples in the neighbor layer $l$ are aggregated from the representations of both the node itself and semantically similar nodes, compensating for the negative impact of arbitrarily treating semantically similar nodes as negative samples in traditional CL. For \textbf{P2}, while nodes with similar preferences share semantics, individual personalization leads to differences that GCNs exploit to enhance representations through neighbor aggregation. These semantic differences can be viewed as beneficial noise in constructing contrastive views, enhancing rather than hindering representation learning.
\section{Methodology}
In this section, we introduce a novel CL-based recommendation paradigm called Neighbor Layers Graph Contrastive Learning (NLGCL). Specifically, we first introduce the naturally existing contrastive views in neighbor layers. Then, we detailed our tailored loss function. Finally, a deeper analysis of our NLGCL is further proposed.

\subsection{Contrastive Views}
We use LightGCN as our graph convolution backbone, obtaining node embeddings $\mathbf{e}_{u}^{(0)}/\mathbf{e}_{i}^{(0)},\mathbf{e}_{u}^{(1)}/\mathbf{e}_{i}^{(1)},...,\mathbf{e}_{u}^{(L)}/\mathbf{e}_{i}^{(L)},$ at each layer, as in Eq~\eqref{eq: neighbor1}. We introduce two scopes of contrastive views: (a) \textbf{heterogeneous} and (b) \textbf{entire}. In the heterogeneous scope, based on the analysis in Section~\ref{sec: ncv}, all item embeddings in layer $(l-1)$ and all user embeddings in layer $l$ form one pair of naturally contrastive views; conversely, all user embeddings in layer $(l-1)$ and all item embeddings in layer $l$ form another pair. In the entire scope, we consider all embeddings in layer $(l-1)$ and all embeddings in layer $l$ as naturally contrastive views. Next, we define the positive and negative pairs within these contrastive views.

\subsubsection{Positive Pairs} Two scopes have the same positive pairs. Given a user $u$, the neighbor set for $u$ is $\mathcal{N}_{u}$. For user $u$, the $(l)$-th layer embeddings of $u$'s neighbors construct the positive pairs with the $(l$$-$$1)$-th layer embedding $\mathbf{e}_{u}^{(l-1)}$ of user $u$. Specifically, $\mathbf{e}_{u}^{(l-1)}$ constructs positive pairs with $\mathbf{e}_{\tilde{i}}^{(l)}$, where $\tilde{i} \in \mathcal{N}_{u}$. The reason for this because $\mathbf{e}_{\tilde{i}}^{(l)}$ are weighted aggregated by $\mathbf{e}_{\bar{u}}^{(l-1)}$, where $\bar{u} \in \mathcal{N}_{\tilde{i}}$ (as Eq~\eqref{eq: neighbor2}). Similar positive pairs are defined for item $i$.

\subsubsection{Negative Pairs} In the heterogeneous scope, the $(l)$-th layer embeddings of all items, excluding items $\tilde{i}$, construct negative pairs with the $(l$$-$$1)$-th layer embedding $\mathbf{e}_{u}^{(l-1)}$ of user $u$, where $\tilde{i} \in \mathcal{N}_{u}$. Similar negative pairs are defined for item $i$. In the entire scope, the $(l)$-th layer embeddings of all users additionally construct negative pairs with the $(l$$-$$1)$-th layer embedding $\mathbf{e}_{u}^{(l-1)}$ of user $u$, while still maintains all negative pairs in the heterogeneous scope. Similar negative pairs are defined for item $i$. We categorize the positive and negative pairs in Table~\ref{tab: cv}.
\begin{table}[h]
    \centering
\caption{Positive and Negative pairs in Contrastive Views.}
\vskip -0.1in
\label{tab: cv}
\resizebox{\linewidth}{!}{
    \begin{tabular}{c|c|cc}
    \hline
        Node & Scope & Positive Pairs & Negative Pairs \\
        \hline
        $u$: $\mathbf{e}_{u}^{(l-1)}$ & Heterogeneous & $\mathbf{e}_{\tilde{i}}^{(l)}$$|$$\tilde{i} \in \mathcal{N}_{u}$ & $\mathbf{e}_{\hat{i}}^{(l)}$$|$$\hat{i} \notin \mathcal{N}_{u}$ \\
        $u$: $\mathbf{e}_{u}^{(l-1)}$ & Entire & $\mathbf{e}_{\tilde{i}}^{(l)}$$|$$\tilde{i} \in \mathcal{N}_{u}$ & $\mathbf{e}_{\hat{i}}^{(l)}$ $\cup$ $\mathbf{e}_{u}^{(l)}$$|$$\hat{i} \notin \mathcal{N}_{u}, u \in \mathcal{U}$\\
        $i$: $\mathbf{e}_{i}^{(l-1)}$ & Heterogeneous & $\mathbf{e}_{\tilde{u}}^{(l)}$$|$$\tilde{u} \in \mathcal{N}_{i}$ & $\mathbf{e}_{\hat{u}}^{(l)}$$|$$\hat{u} \notin \mathcal{N}_{i}$ \\
        $i$: $\mathbf{e}_{i}^{(l-1)}$ & Entire & $\mathbf{e}_{\tilde{u}}^{(l)}$$|$$\tilde{u} \in \mathcal{N}_{i}$ & $\mathbf{e}_{\hat{u}}^{(l)}$ $\cup$ $\mathbf{e}_{i}^{(l)}$$|$$\hat{u} \notin \mathcal{N}_{i}, i \in \mathcal{I}$\\
        \hline
    \end{tabular}
    }
\end{table}

\textbf{\textit{Analysis}}:
Unlike traditional contrastive learning, our approach associates each node with multiple positive samples, where noise among positives arises from semantically similar nodes. This enhances the alignment of positive samples while reducing the impact of random noise. Additionally, it addresses the limitation of traditional CL, which arbitrarily treats semantically similar nodes as negatives. Both scopes use the same positive sample selection, but the entire scope provides a broader range of negatives. While a larger negative sample scope can improve positive feature learning, it may hinder semantic alignment between users and items and increase computational costs. We empirically validate the effectiveness and efficiency of both scopes in Sections~\ref{sec: scope} and~\ref{subsec: efficiency}.

\subsection{Contrastive Learning Loss}
As discussed in Section~\ref{sec: 3B}, traditional CL considers a node and its counterpart in another view as positive pairs, while treating the node and all other nodes in different views as negative pairs. However, having a small number of positive pairs and arbitrarily defined negative pairs can irrationally push nodes with similar semantics farther away. In contrast, our tailored CL method assigns each node a set of positive pairs, enhancing alignment and reducing the impact of random noise. In an $L$-layer GNN, up to $L$ groups of contrastive views can be constructed, but this incurs additional computational overhead. Let $G$ denote the number of contrastive view groups, where $G \leq L$. Selecting these $G$ groups is crucial, and we theoretically prove in Theorem~\ref{th:1} (Appendix~\ref{appendix: proof}) that the first $G$ groups are optimal.

\begin{theorem}
\label{th:1}
In GCNs, contrastive views that naturally exist between layer $(l)$ and layer $(l$$+$$1)$ are more effective for contrastive learning when $(l)$ is smaller.
\end{theorem}

Based on Theorem~\ref{th:1}, selecting the first $G$ layers and their neighbor layers as contrastive views achieves optimal results. We introduce our tailored CL loss for two scopes of contrastive views, respectively.

\subsubsection{Heterogeneous Scope} For the heterogeneous scope, our neighbor layer CL loss can be formulated as:

\vskip -0.25in
\begin{equation}
\resizebox{0.9\hsize}{!}{$\begin{aligned}
        \mathcal{L}_{nl_u} = - \frac{1}{G|\mathcal{U}|} \sum_{g = 0}^{G-1}\sum_{u \in \mathcal{U}} \frac{1}{|\mathcal{N}_u|}  \log \frac{\prod\limits_{i^{+} \in \mathcal{N}_u} \exp({{{\mathbf{e}_u^{(g)}}^{\top}\mathbf{e}_{i^+}^{(g+1)}}/{\tau}})}{\sum\limits_{\hat{i} \in \mathcal{I}} \exp({{\mathbf{e}_u^{(g)}}^{\top}\mathbf{e}_{\hat{i}}^{(g+1)}/\tau})},
\end{aligned}$}
\end{equation}
\vskip -0.15in
\begin{equation}
\resizebox{0.9\hsize}{!}{$\begin{aligned}
    \mathcal{L}_{nl_i} = - \frac{1}{G|\mathcal{I}|} \sum_{g = 0}^{G-1}\sum_{i \in \mathcal{I}} \frac{1}{|\mathcal{N}_i|} \log \frac{\prod\limits_{u^{+} \in \mathcal{N}_i} \exp({{\mathbf{e}_i^{(g)}}^{\top}\mathbf{e}_{u^+}^{(g+1)}/{\tau}})}{\sum\limits_{\hat{u} \in \mathcal{U}} \exp({{{\mathbf{e}_i^{(g)}}^{\top}\mathbf{e}_{\hat{u}}^{(g+1)}}/{\tau}})},
\end{aligned}$}
\end{equation}
\vskip -0.05in
\noindent where positive samples $i^{+}/u^{+}$ refer to all neighbors of each $u/i$, while negative samples $\hat{i}/\hat{u}$ are drawn from the entire set of $\mathcal{I}$ or $\mathcal{U}$. We compute the average over the first $G$ layers for all users/items.

\subsubsection{Entire Scope} For the entire scope, our neighbor layer CL loss can be formulated as:
\vskip -0.15in
\begin{equation}
\resizebox{0.9\hsize}{!}{$\begin{aligned}
    \mathcal{L}_{nl_u} = - \frac{1}{G|\mathcal{U}|} \sum_{g = 0}^{G-1}\sum_{u \in \mathcal{U}} \frac{1}{|\mathcal{N}_u|}  \log \frac{\prod\limits_{i^{+} \in \mathcal{N}_u}\exp({{{\mathbf{e}_u^{(g)}}^{\top}\mathbf{e}_{i^+}^{(g+1)}}/{\tau}})}{\sum\limits_{\hat{x} \in \mathcal{V}} \exp({{{\mathbf{e}_u^{(g)}}^{\top}\mathbf{e}_{\hat{x}}^{(g+1)}}/{\tau}})},
\end{aligned}$}
\end{equation}
\vskip -0.15in
\begin{equation}
\resizebox{0.9\hsize}{!}{$\begin{aligned}
    \mathcal{L}_{nl_i} = - \frac{1}{G|\mathcal{I}|} \sum_{g = 0}^{G-1}\sum_{i \in \mathcal{I}} \frac{1}{|\mathcal{N}_i|}  \log \frac{\prod\limits_{u^{+} \in \mathcal{N}_i}\exp({{{\mathbf{e}_i^{(g)}}^{\top}\mathbf{e}_{u^+}^{(g+1)}}/{\tau}})}{\sum\limits_{\hat{x} \in \mathcal{V}} \exp({{{\mathbf{e}_i^{(g)}}^{\top}\mathbf{e}_{\hat{x}}^{(g+1)}}/{\tau}})},
\end{aligned}$}
\end{equation}
\vskip -0.05in
\noindent where positive samples $i^{+}/u^{+}$ refer to all neighbors of each $u/i$, while negative samples $\hat{x}$ are drawn from the entire set $\mathcal{V} = \mathcal{I} \cup \mathcal{U}$. We compute the average over the first $G$ layers for all users/items. 

For any scope, we get the final loss $\mathcal{L}_{nl}$ by summing both user-side loss $\mathcal{L}_{nl_u}$ and item-side loss $\mathcal{L}_{nl_i}$, formally: $ \mathcal{L}_{nl} = \mathcal{L}_{nl_u} + \mathcal{L}_{nl_i}$.

\textbf{\textit{Efficiency:}} Although our NLGCL introduces a computational complexity proportional to $G$, the cost of constructing contrastive views in traditional CL is significantly higher than the cost of computing the CL loss itself. By leveraging the naturally existing contrastive views within GCNs, our NLGCL eliminates the substantial overhead of view construction. Detailed analysis of time efficiency is provided in Appendices~\ref{appendix: analysis efficiency}.

\subsection{Model Optimization}
To extract node representations in NLGCL, we adopt a multi-task training strategy that jointly optimizes the traditional BPR loss (Eq~\eqref{eq: bpr}) and our neighbor layers graph contrastive learning loss:
\begin{equation}
\mathcal{L} = \mathcal{L}_{bpr} + \lambda_1 \mathcal{L}_{nl} + \lambda_2 \|\mathbf{\Theta}\|^2,
\end{equation}
where $\lambda_1$ and $\lambda_2$ are balancing hyper-parameters of CL and $L_2$ regularization term, respectively. $\mathbf{\Theta}$ denotes model parameters. 

\begin{table}[!ht]
    \centering
    \small
\caption{Statistics of four datasets.}
 \vskip -0.1in
\label{tab: dataset}
    \begin{tabular}{ccccc}
     \hline
         Dataset&  \# Users&  \# Items&  \# Interaction& Sparsity\\
         \hline
         Yelp & 45,477 & 30,708 & 1,777,765 & 99.873\%\\
         Pinterest & 55,188 & 9,912 & 1,445,622 & 99.736\%\\
         QB-Video & 30,324 & 25,731 & 1,581,136 & 99.797\%\\
         Alibaba & 300,001 & 81,615 & 1,607,813 & 99.993\%\\
        \hline
    \end{tabular}
\end{table}

\section{Experiments}
\label{sec: experiments}

In this section, we briefly describe our experimental settings and then conduct extensive experiments on four public datasets to evaluate our proposed NLGCL by answering the following research questions: \textbf{RQ1:} How does our proposed NLGCL perform in comparison to various state-of-the-art recommender systems? \textbf{RQ2:} How do the different scopes of our NLGCL impact its performance? \textbf{RQ3:} How efficient is our NLGCL compared with various state-of-the-art recommender systems? \textbf{RQ4:} Can our NLGCL learn high-quality representations with uniform distribution? \textbf{RQ5:} How do different hyper-parameters influence?
\begin{table*}[!ht]
    \centering
    \small
\caption{Performance comparison of baselines and our NLGCL in terms of Recall@K(R@K) and NDCG@K(N@K). The superscript $^*$ indicates the improvement is statistically significant where the $p$-value is less than 0.01.}
\vskip -0.1in
\label{tab: comparison results}
\resizebox{\linewidth}{!}{
    \begin{tabular}{cc|ccccc|cccccc|cc}
    \hline
     & Model & MF-BPR & NGCF & LightGCN & IMP-GCN & LayerGCN & SGL & NCL & SimGCL & LightGCL & DCCF & BIGCF & NLGCL & \multirow{2}{*}{Improv.}\\ \cline{2-14}
      & Metric & \cite{rendle2009bpr} & \cite{wang2019neural} & \cite{he2020lightgcn} & \cite{liu2021interest} & \cite{zhou2023layer}  & \cite{wu2021self} & \cite{lin2022improving} & \cite{yu2022graph} & \cite{cailightgcl} & \cite{ren2023disentangled} & \cite{zhang2024exploring} & Our& \\ \hline
  
     \multirow{6}{*}{Yelp} 
    & R@10 & 0.0643& 0.0630& 0.0730& 0.0751& 0.0771
    & 0.0833& 0.0902& \underline{0.0908}& 0.0766& 0.0818& 0.0880 & \textbf{0.0952$^*$} & 4.85\%\\
    & R@20 & 0.1043& 0.1026& 0.1163& 0.1182& 0.1208
    & 0.1288& 0.1325& \underline{0.1331}& 0.1188& 0.1268& 0.1311 & \textbf{0.1414$^*$} & 6.24\%\\
    & R@50 & 0.1862& 0.1864& 0.2016& 0.2017& 0.2041
    & 0.2100& 0.2127& \underline{0.2130}& 0.2008& 0.2107& 0.2119 & \textbf{0.2327$^*$} & 9.25\%\\ \cline{2-15}
    & N@10& 0.0458& 0.0446& 0.0520& 0.0539& 0.0560
    & 0.0601& 0.0673& \underline{0.0682}& 0.0551& 0.0596& 0.0630 & \textbf{0.0713$^*$} & 4.55\%\\
    & N@20& 0.0580& 0.0567& 0.0652& 0.0662& 0.0684
    & 0.0739& 0.0811& \underline{0.0823}& 0.0681& 0.0741& 0.0779 & \textbf{0.0859$^*$} & 4.37\%\\
    & N@50& 0.0793& 0.0784& 0.0875& 0.0885& 0.0901
    & 0.0964& 0.1033& \underline{0.1039}& 0.0899& 0.0974& 0.1005 & \textbf{0.1094$^*$} & 5.29\%\\ \hline
    \multirow{6}{*}{Pinterest}
    & R@10 & 0.0855& 0.0870& 0.1000& 0.0985& 0.1004
    & \underline{0.1080}& 0.1033& 0.1051& 0.0881& 0.1040& 0.1040& \textbf{0.1148$^*$} & 6.30\%\\
    & R@20 & 0.1409& 0.1428& 0.1621& 0.1603& 0.1620
    & \underline{0.1704}& 0.1609& 0.1576& 0.1322& 0.1613& 0.1619& \textbf{0.1793$^*$} & 5.22\%\\
    & R@50 & 0.2599& 0.2633& 0.2862& 0.2845& 0.2880
    & \underline{0.2963}& 0.2887& 0.2442& 0.2383& 0.2871& 0.2894& \textbf{0.3089$^*$} & 4.25\%\\ \cline{2-15}
    & N@10& 0.0537& 0.0545& 0.0635& 0.0624& 0.0635
    & 0.0701& 0.0666& \underline{0.0705}& 0.0534& 0.0661& 0.0680& \textbf{0.0760$^*$} & 7.80\%\\
    & N@20& 0.0708& 0.0721& 0.0830& 0.0814& 0.0826
    & \underline{0.0897}& 0.0833& 0.0871& 0.0673& 0.0828& 0.0864& \textbf{0.0948$^*$} & 5.69\%\\
    & N@50 & 0.1001& 0.1018& 0.1136& 0.1113& 0.1121
    & \underline{0.1209}& 0.1150& 0.1086& 0.0981& 0.1157& 0.1155& \textbf{0.1267$^*$} & 4.80\%\\ \hline
    \multirow{6}{*}{QB-Video} 
    & R@10 & 0.1120& 0.1261& 0.1275& 0.1278& 0.1291
    & 0.1341& \underline{0.1372}& 0.1337& 0.1330& 0.1350& 0.1341& \textbf{0.1427$^*$} & 4.01\%\\
    & R@20 & 0.1741& 0.1943& 0.1969& 0.1966& 0.1990
    & 0.2030& \underline{0.2089}& 0.2030& 0.2018& 0.2044& 0.2056& \textbf{0.2167$^*$} & 3.73\%\\
    & R@50 & 0.2969& 0.3237& 0.3251& 0.3259& 0.3271
    & 0.3342& 0.3414& 0.3356& 0.3258& 0.3281& \underline{0.3423}& \textbf{0.3501$^*$} & 2.28\%\\ \cline{2-15}
    & N@10 & 0.0782& 0.0888& 0.0900& 0.0899& 0.0904
    & 0.0932& \underline{0.0964}& 0.0945& 0.0921& 0.0951& 0.0928& \textbf{0.0996$^*$} & 3.32\%\\
    & N@20 & 0.0977& 0.1099& 0.1109& 0.1105& 0.1123
    & 0.1141& \underline{0.1189}& 0.1151& 0.1130& 0.1150& 0.1144& \textbf{0.1219$^*$} & 2.52\%\\
    & N@50& 0.1303& 0.1438& 0.1448& 0.1447& 0.1461
    & 0.1482& \underline{0.1530}& 0.1501& 0.1459& 0.1508& 0.1503& \textbf{0.1573$^*$} & 2.81\%\\ \hline
     \multirow{6}{*}{Alibaba} 
    & R@10 & 0.0303& 0.0382& 0.0457& 0.0400& 0.0448
    & 0.0461& 0.0477& 0.0474& 0.0459& 0.0490& \underline{0.0502}& \textbf{0.0531$^*$} & 5.78\%\\
    & R@20 & 0.0467& 0.0615& 0.0692& 0.0635& 0.0680
    & 0.0692& 0.0713& 0.0691& 0.0716& 0.0729& \underline{0.0744}& \textbf{0.0773$^*$} & 3.90\%\\
    & R@50 & 0.0799& 0.1081& 0.1144& 0.1110& 0.1138
    & 0.1141& 0.1165& 0.1092& 0.1204& 0.1199& \underline{0.1205}& \textbf{0.1238$^*$} & 2.74\%\\ \cline{2-15}
    & N@10& 0.0161& 0.0198& 0.0246& 0.0221& 0.0238
    & 0.0248& 0.0259& 0.0262& 0.0239& 0.0257& \underline{0.0266}& \textbf{0.0299$^*$} & 12.41\%\\
    & N@20& 0.0203& 0.0257& 0.0299& 0.0271& 0.0285
    & 0.0307& 0.0319& 0.0317& 0.0305& 0.0311& \underline{0.0322}& \textbf{0.0361$^*$} & 12.11\%\\
    & N@50& 0.0269& 0.0349& 0.0396& 0.0369& 0.0393
    & 0.0396& 0.0409& 0.0397& 0.0410& 0.0404& \underline{0.0415}& \textbf{0.0450$^*$} & 8.43\%\\ \hline
    \end{tabular}
}
\end{table*}

\begin{table*}[!ht]
\centering
\small
\tabcolsep=0.1cm
\caption{Efficiency comparison of different methods across four datasets, including average training time per epoch, number of epochs to converge, total training time (T/E: Time/Epoch, \#E: \#Epoch, TT: Total Time, N@10: NDCG@10; s: second, m: minute, h: hour). We highlight the optimal and suboptimal models for the TT metric in \textbf{bold} and \underline{underline}, respectively.}
 \vskip -0.1in
\label{tab: efficiency}
\resizebox{1\linewidth}{!}{
\begin{tabular}{ccccccccccccccccc}
\hline
\multirow{2.5}{*}{\textbf{Model}}&\multicolumn{4}{c}{\textbf{Yelp}} & \multicolumn{4}{c}{\textbf{Pinterest}} & \multicolumn{4}{c}{\textbf{QB-Video}} & \multicolumn{4}{c}{\textbf{Alibaba}}\\ \cline{2-5} \cline{6-9} \cline{10-13} \cline{14-17}& T/E$\downarrow$ & {\#E$\downarrow$} & TT$\downarrow$ & {N@10$\uparrow$} & T/E$\downarrow$ & {\#E$\downarrow$} & TT$\downarrow$ & {N@10$\uparrow$} & T/E$\downarrow$ & {\#E$\downarrow$} & TT$\downarrow$ & {N@10$\uparrow$} & T/E$\downarrow$ & {\#E$\downarrow$} & TT$\downarrow$ & {N@10$\uparrow$} \\
\hline
LightGCN & 22.42s& 332& 2h9m& 0.0520& 8.40s& 194& 27m& 0.0635& 25.32s& 283& 1h59m& 0.0900& 21.55s& 371& 2h13m& 0.0246\\ \hline
SGL & 104.19s& 55& 1h36m& 0.0601& 42.71s& 43& 31m& 0.0701& 124.44s& 61& 2h7m& 0.0932& 95.37s& 83& 2h12m& 0.0248\\
NCL & 94.42s& 102& 2h41m& 0.0673& 35.10s& 83& 49m& 0.0666& 108.98s& 90& 2h43m& 0.0964& 85.79s& 88& 2h6m& 0.0477\\
SimGCL & 98.33s& 155& 4h14m& 0.0682& 37.31s& 121& 1h15m& 0.0705& 110.92s& 167& 5h9m& 0.0945& 87.31s& 241& 5h51m& 0.0262\\
LightGCL & 100.51s& 64& 1h47m& 0.0551& 38.87s& 51& 33m& 0.0534& 119.18s& 57& 1h53m& 0.0921& 92.24s& 69& 1h46m& 0.0459\\
DCCF & 221.31s& 38& 2h20m& 0.0596& 76.39s& 32& 41m& 0.0661& 228.21s& 38& 2h25m& 0.0951& 180.08s& 49& 2h27m& 0.0257\\
BIGCF & 208.19s& 41& 2h22m& 0.0630& 70.11s& 35& 41m& 0.0680& 206.60s& 38& 2h11m& 0.0928& 167.12s& 42& 1h57m& 0.0266\\ \hline
NLGCL-H & 46.93s& 63& \textbf{49m}& 0.0713& 19.55s& 40& \textbf{13m}& 0.0760& 53.19s& 55& \textbf{49m}& 0.0996& 44.73s& 74& \textbf{55m}& 0.0299\\
NLGCL-E & 61.32s& 82& \underline{1h28m}& 0.0689& 28.81s& 51& \underline{24m}& 0.0729& 72.99s& 62& \underline{1h15m}& 0.0969& 59.67s& 79& \underline{1h19m}& 0.0279\\ \hline

\end{tabular}
}
\end{table*}

\subsection{Experimental Settings}

\subsubsection{Datasets}
To validate the effectiveness of NLGCL, we conduct experiments on four public datasets, namely \textbf{Yelp} \cite{lin2022improving}, \textbf{Pinterest} \cite{geng2015learning}, \textbf{QB-Video} \cite{yuan2022tenrec}, and \textbf{Alibaba} \cite{chen2019pog}. These datasets originate from diverse domains, exhibiting variances in both scale and sparsity. To ensure the quality of the data, we employ the 15-core setting for the \textbf{Yelp} and \textbf{Alibaba} datasets, which ensures a minimum of 15 interactions between users and items. For the \textbf{Pinterest} and \textbf{QB-Video} datasets, users and items with less than 5 interactions are filtered out. Table~\ref{tab: dataset} summarizes statistical information for all datasets. We use the user-based split approach to divide the data for experimentation. Specifically, we divide training, validation, and test sets into a ratio of 8:1:1, respectively. Moreover, we employ pair-wise sampling, where a negative interaction is randomly chosen for each interaction in the training set to construct the training sample.

\subsubsection{Metrics} To evaluate the top-$K$ recommendation task performance fairly, we adopt two widely-used metrics: Recall and Normalized Discounted Cumulative Gain (NDCG). The values of $K$ are set to {10, 20, 50}. Following previous work \cite{lin2022improving}, we adopt the all-rank protocol which ranks all candidate items that users have not interacted with during evaluation.

\subsubsection{Implementation Details}
To ensure a fair comparison, we implement our NLGCL\footnote[2]{Code is available at: \href{https://github.com/Jinfeng-Xu/NLGCL}{https://github.com/Jinfeng-Xu/NLGCL}.} and all baselines using the RecBole and RecBole-GNN \cite{zhao2022recbole}, unified frameworks for traditional recommendation models and graph-based traditional recommendation models, respectively. We use the Adam optimizer \cite{kingma2014adam} and Xavier initialization \cite{glorot2010understanding} with default parameters, and conduct extensive hyper-parameter tuning for all baselines. For our NLGCL, we fix the batch size to 4096, set $\lambda_2$ for $L_2$ regularization to $10^{-4}$, and use an embedding size of 64. Early stopping is applied with an epoch limit of 20, monitored by NDCG@10. For our NLGCL, we tune hyper-parameters $\lambda_1 \in \{10^{-6}, 10^{-5}, 10^{-4}\}$, temperature $\tau \in \{0.1, 0.2, 0.3, 0.4\}$, the number of GCN layers $L \in \{1, 2, 3, 4\}$, and the number of contrastive view groups $G \in \{1, ..., L\}$.

\subsubsection{Baselines}
To assess the effectiveness of NLGCL, we compare it against two categories of baselines: (1) CF-based models (\textbf{MF-BPR} \cite{rendle2009bpr}, \textbf{NGCF} \cite{wang2019neural}, \textbf{LightGCN} \cite{he2020lightgcn}, \textbf{IMP-GCN} \cite{liu2021interest}, and \textbf{LayerGCN} \cite{zhou2023layer}) and (2) CL-based models (\textbf{SGL} \cite{wu2021self}, \textbf{NCL} \cite{lin2022improving}, \textbf{SimGCL} \cite{yu2022graph}, \textbf{LightGCL} \cite{cailightgcl}, \textbf{DCCF} \cite{ren2023disentangled}, and \textbf{BIGCF} \cite{zhang2024exploring}).

\subsection{Overall Performance (RQ1)}
\label{subsec: effectiveness}
Table~\ref{tab: comparison results} reports the recommendation performance of all baselines on four public datasets. From the results, we observed that:

\noindent \textbf{Observation1:} Our proposed NLGCL achieves the best performance, outperforming all baselines and demonstrating its effectiveness in recommendation tasks. Specifically, NLGCL improves NDCG@10 by 4.55\%, 5.69\%, 2.52\%, and 12.11\% on Yelp, Pinterest, QB-Video, and Alibaba, respectively, compared to the strongest baseline. These results highlight NLGCL's strong capability in providing personalized recommendations.

\noindent \textbf{Observation2:} Across all datasets, CL-based methods generally outperform CF-based methods, highlighting the benefits of integrating contrastive learning with collaborative filtering. However, existing CL-based methods often lose crucial information and introduce noise during view construction. In contrast, NLGCL mitigates this issue by associating multiple positive samples with each node, where noise among these samples reflects other nodes with similar preferences.
%
\subsection{The Influence of Different Scopes in Contrastive Views (RQ2)}
\label{sec: scope}
\begin{figure}
    \centering
    \includegraphics[width=1\linewidth]{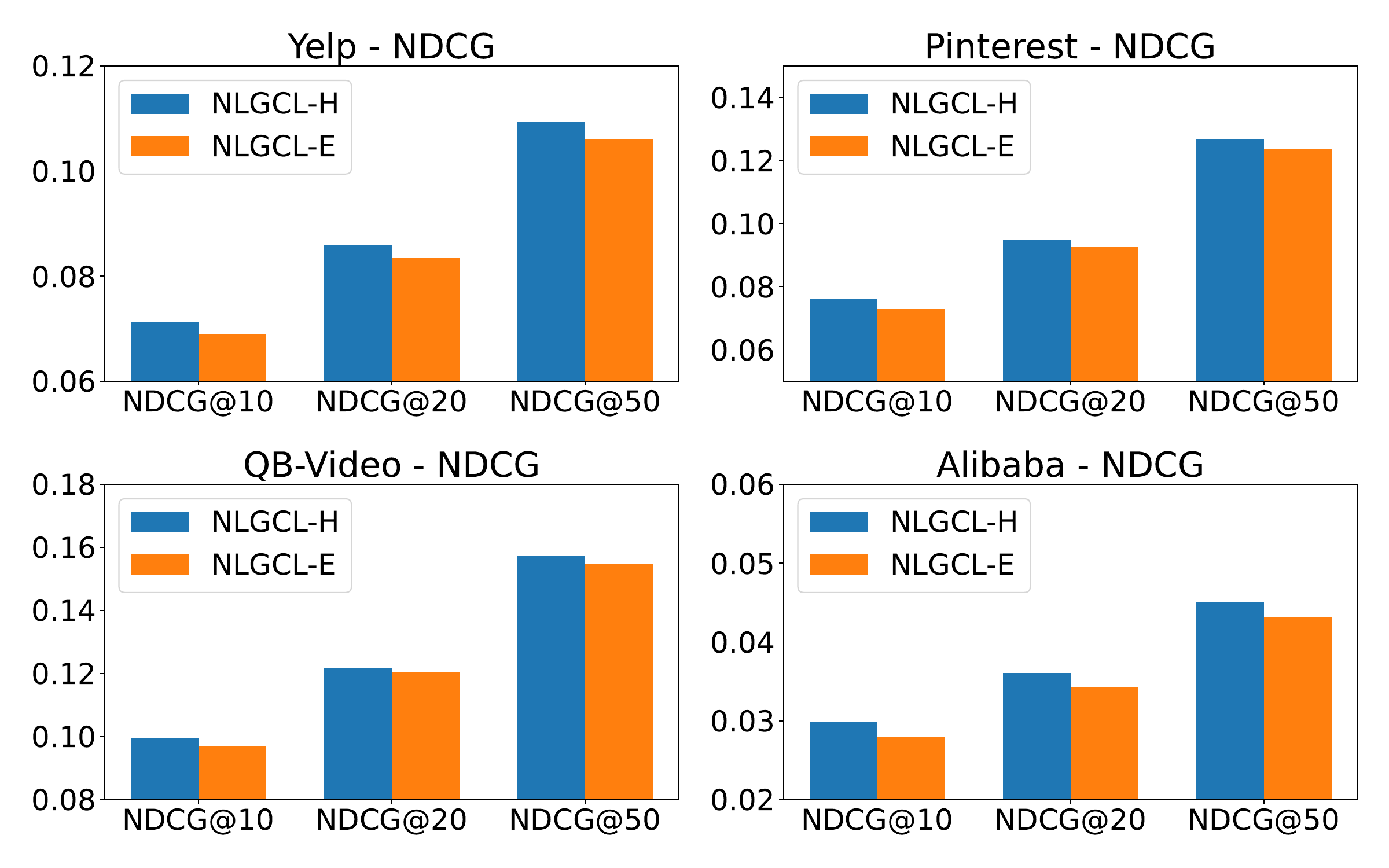}
    \caption{The influence of the scope of contrastive views.}
    \label{fig: scope}
\end{figure}

To validate the effectiveness of different scopes of contrastive views, we conduct experiments on all datasets. We design the following variants: 1) NLGCL-H, which adopts the heterogeneous scope. 2) NLGCL-E, which adopts the entire scope. In Figure~\ref{fig: scope}, we observed that NLGCL-H consistently outperforms NLGCL-E across all datasets. This advantage is attributable to the scope of NLGCL-H being more accurate than NLGCL-E. Specifically, there is an inevitable semantic gap between users and items. To this end, arbitrarily treating all nodes as negative samples will aggravate semantic noise in user/item representation and hurt recommendation accuracy. Moreover, NLGCL-E also requires higher computational resources than NLGCL-H, which we detail in Section~\ref{subsec: efficiency}.

    

\begin{figure}[!h]
    \centering
    \includegraphics[width=1\linewidth]{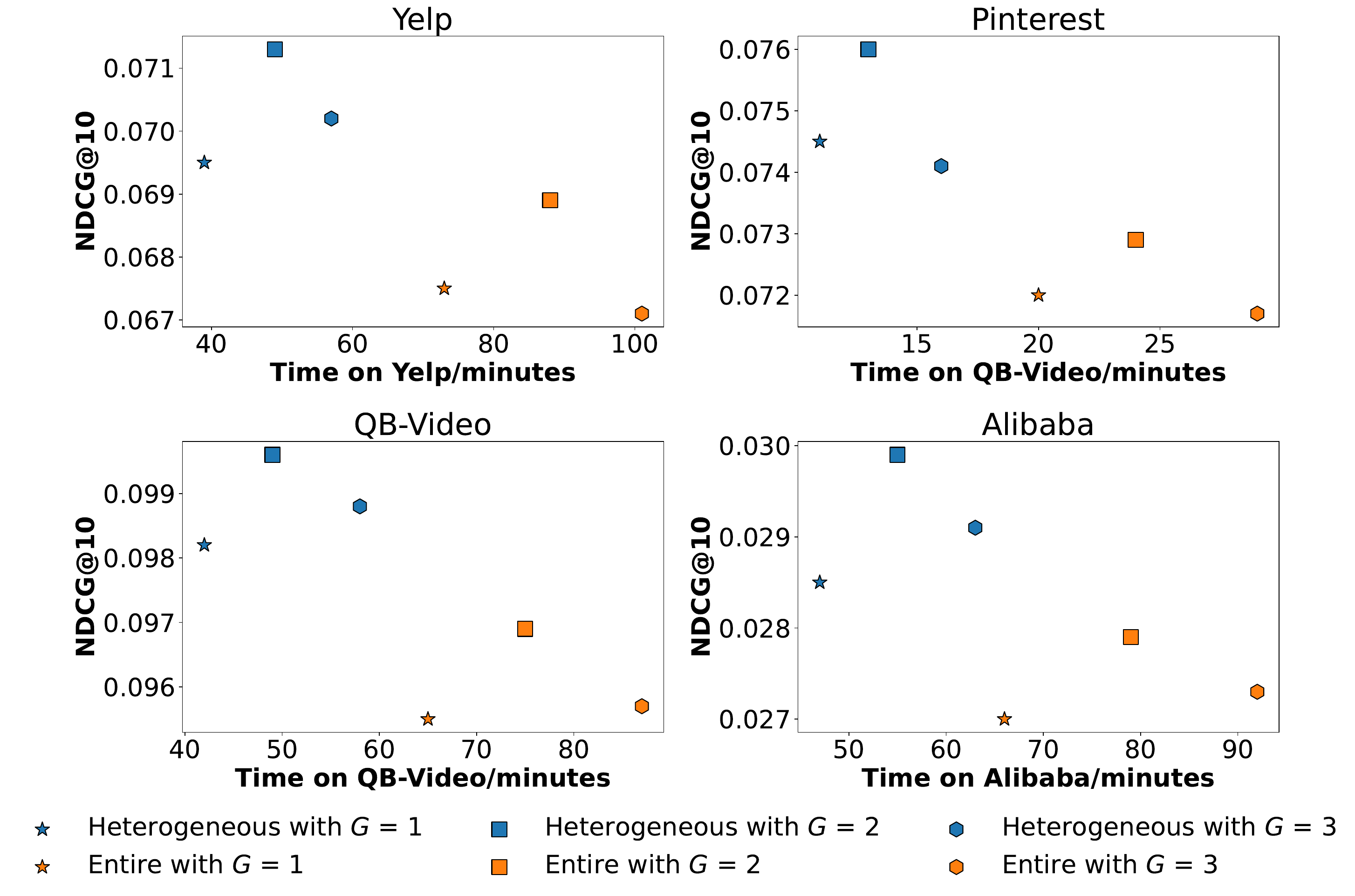}
    \caption{Efficiency study in terms of NDCG@10.}
    \label{fig: G}
\end{figure}

\begin{figure*}[!t]
    \centering
    \includegraphics[width=1\linewidth]{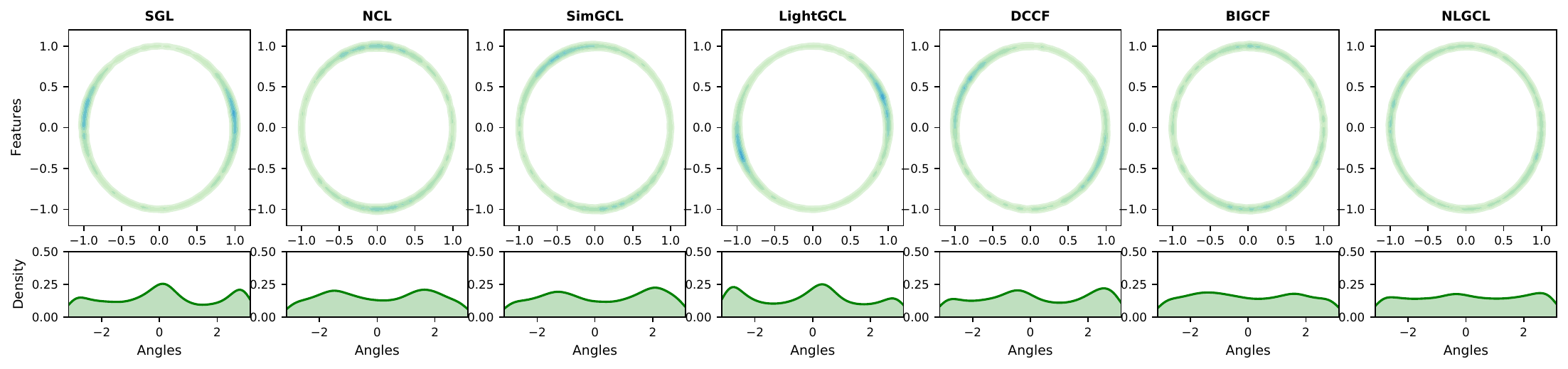}
    \caption{We follow previous work \cite{yu2022graph} to plot feature distributions with Gaussian kernel density estimation (KDE) in $\mathcal{R}^2$ (the darker the color is, the more points fall in that area.) and KDE on angles (i.e., arctan2(y, x) for each point (x,y)).}
    \label{fig: vis}
\end{figure*}


\begin{figure*}[!t]
    \centering
    \subfigure {
        \label{fig:Heatmap_Yelp}
        \includegraphics[width=0.23\linewidth]{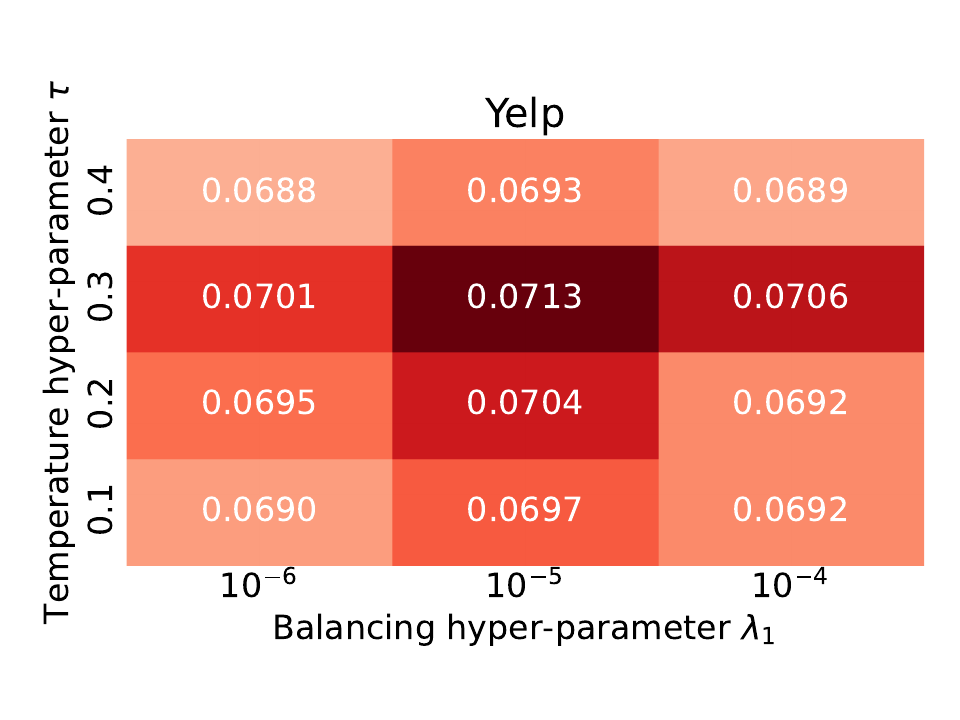}
        } 
    \subfigure {
        \label{fig:Heatmap_Pinterest}
        \includegraphics[width=0.23\linewidth]{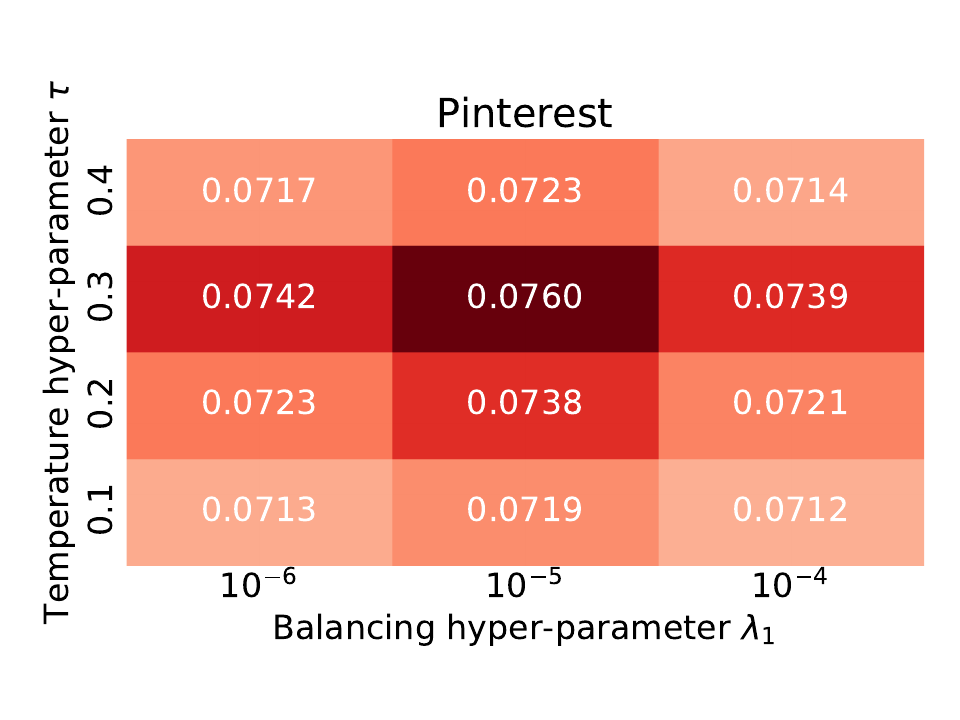}
        }
    \subfigure {
        \label{fig:Heatmap_QB-Video}
        \includegraphics[width=0.23\linewidth]{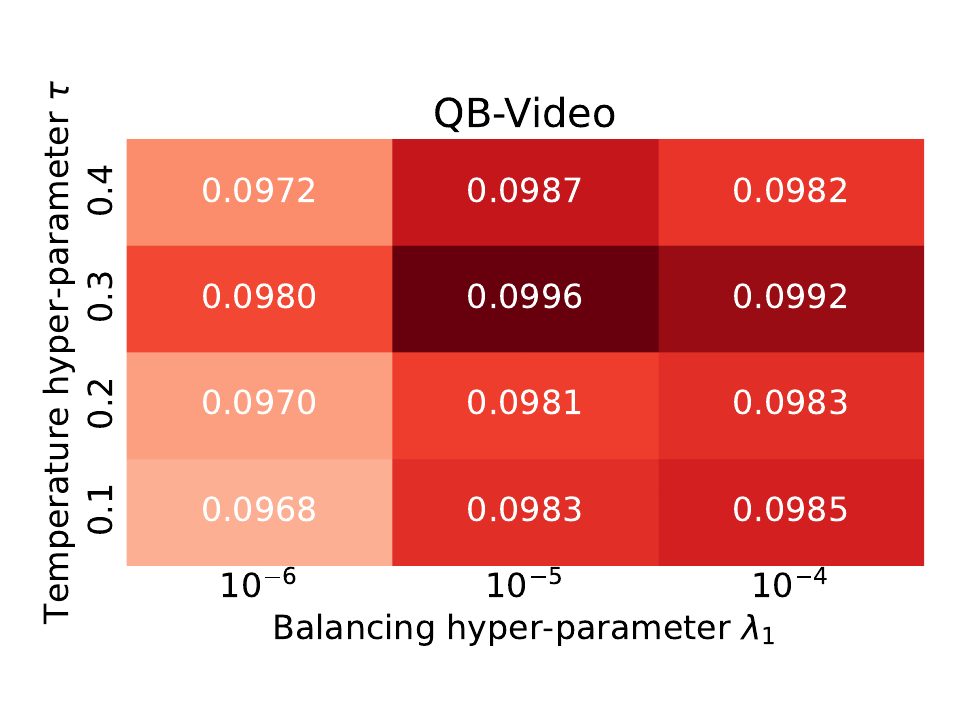}
        }
    \subfigure {
        \label{fig:Heatmap_Alibaba}
        \includegraphics[width=0.23\linewidth]{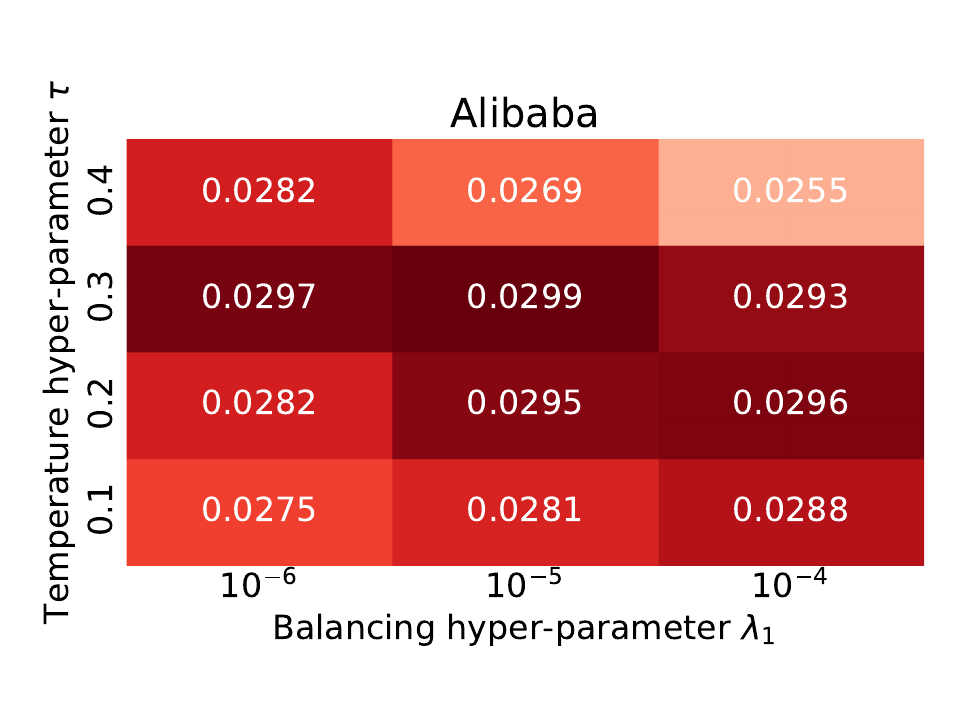}
        }
        \vskip -0.2in
    \caption{Performance Comparison $w.r.t.$ $\lambda_1$ and $\tau$.}
    \label{fig: Heatmap for lambda_1 and tau.}
\end{figure*}

\begin{figure*}[!t]
    \centering
    \subfigure {
        \label{fig:Heatmap_Yelp}
        \includegraphics[width=0.23\linewidth]{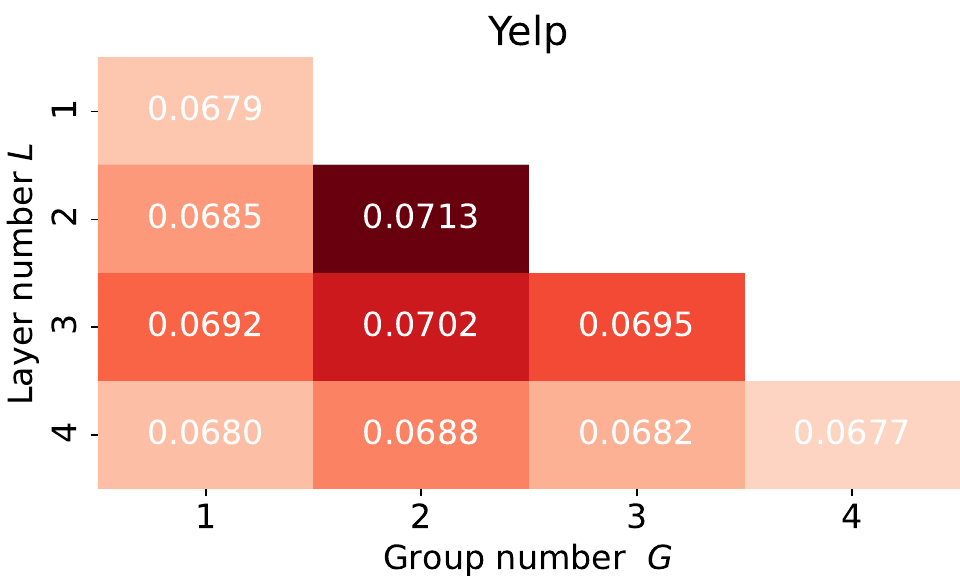}
        }
    \subfigure {
        \label{fig:Heatmap_Pinterest}
        \includegraphics[width=0.23\linewidth]{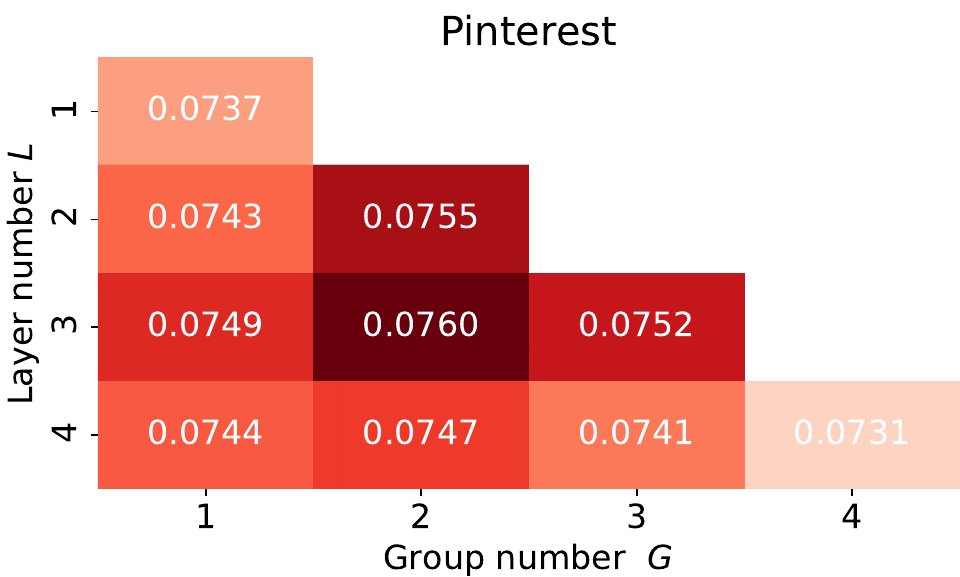}
        }
    \subfigure {
        \label{fig:Heatmap_QB-Video}
        \includegraphics[width=0.23\linewidth]{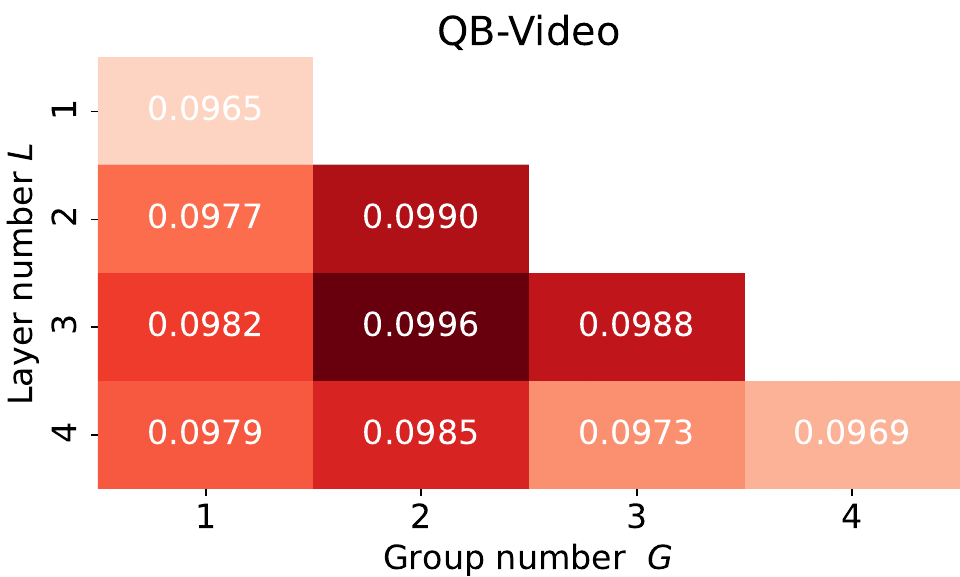}
        }
    \subfigure {
        \label{fig:Heatmap_Alibaba}
        \includegraphics[width=0.23\linewidth]{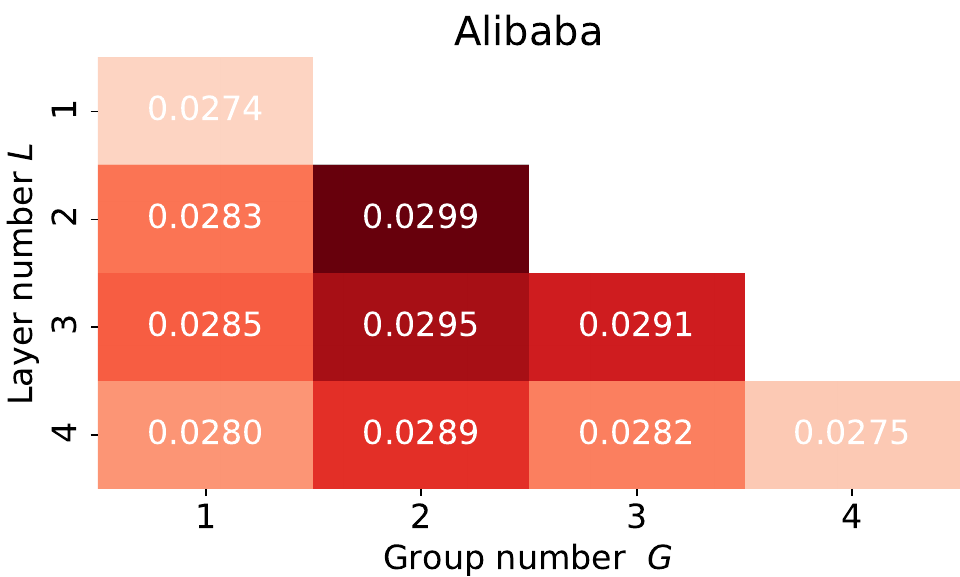}
        }
         \vskip -0.2in
    \caption{Performance Comparison $w.r.t.$ $L$ and $G$.}
    \label{fig: Heatmap for L and G.}
\end{figure*}

\subsection{Efficiency Study (RQ3)}
\label{subsec: efficiency}
We provide theoretical analysis for the efficiency of our proposed NLGCL in Appendix~\ref{appendix: analysis efficiency}. In this section, we further validate it with an empirical study. We conduct experiments across all CL-based baselines using four datasets. In Table~\ref{tab: efficiency}, we present the efficiency of NLGCL and CL-based baselines in terms of average training time per epoch, the number of epochs to converge, and total training time. Our results show that NLGCL achieves competitive per-epoch training efficiency and the fastest convergence among all models. This indicates that NLGCL not only reduces computational time but also reaches optimal performance swiftly, making it ideal for scenarios with limited time or computational resources, such as rapid deployment and frequent updates. As analyzed in Appendix~\ref{appendix: analysis efficiency}, our NLGCL-H and NLGCL-E further reduce training time by adjusting hyper-parameter $G$. Figure~\ref{fig: G} illustrates the trade-off between performance and efficiency across different values of $G$. By tuning $G$, NLGCL adapts to diverse computational environments, balancing efficiency and effectiveness. This flexibility ensures its suitability for real-world applications, even in resource-limited scenarios, while maintaining strong performance.

\subsection{Visualization Analysis (RQ4)}
We validate the effectiveness of NLGCL in learning high-quality, uniformly distributed representations, thereby maximizing the benefits of CL. Specifically, we randomly sample 3,000 users from the Yelp and Pinterest datasets and project their final embeddings from the optimal model onto 2-dimensional unit vectors using t-SNE \cite{van2008visualizing}. As shown in Figure~\ref{fig: vis}, we plot the feature distributions and estimate their angular densities using Gaussian Kernel Density Estimation (KDE) \cite{chen2017tutorial}. Figure~\ref{fig: vis} demonstrates that the representations learned by NLGCL are more uniformly distributed. Consistent with prior studies \cite{zhang2024exploring,yu2022graph,wang2022towards,wang2020understanding}, such uniform distribution enhances the intrinsic quality of representations and maximizes information retention, leading to a better ability to preserve unique user preferences.

\subsection{Hyper-parameter Study (RQ5)}
This section investigates the sensitivity of hyper-parameters on the recommendation performance of NLGCL. The evaluation results in terms of NDCG@10 are reported in Figure~\ref{fig: Heatmap for lambda_1 and tau.} and Figure~\ref{fig: Heatmap for L and G.}.

\noindent\textbf{Performance Comparison $w.r.t.$ $\lambda_1$ and $\tau$:} We analyze the hyper-parameter sensitivity for contrastive learning by varying the balancing hyper-parameter $\lambda_1$ from 10$^{-6}$ to 10$^{-4}$ and temperature parameter $\tau$ from 0.1 to 0.4. As shown in Figure~\ref{fig: Heatmap for lambda_1 and tau.}, NLGCL consistently achieves the best performance with $\lambda_1$ = 10$^{-5}$ and $\tau$ = 0.2 across all datasets, aligning with hyper-parameter settings in traditional CL.

\noindent\textbf{Performance Comparison $w.r.t.$ $L$ and $G$:} We analyze the influence of GCN layer number $L$ and contrastive view group number $G$ across all datasets. Figure~\ref{fig: Heatmap for L and G.} shows that $G$ = 2 is optimal across all datasets, while the optimal $L$ is 2 for Yelp and Alibaba datasets, and 3 for Pinterest and QB-Video datasets. Notably, even with $G$ = 1, NLGCL maintains strong performance, consistent with the efficiency analysis in Section~\ref{subsec: efficiency}. 
\section{Related Work}
\subsection{GNN-based Recommendation}
Graph Neural Networks (GNNs) are widely applied in recommendation systems due to their ability to capture high-order relational information in user-item bipartite graphs. NGCF \cite{wang2019neural} pioneered the use of GNNs for aggregating high-order neighborhood information, laying the foundation for subsequent advancements. GCCF \cite{chen2020revisiting} and LightGCN \cite{he2020lightgcn} enhanced encoding quality by removing non-linear transformations during propagation, which has since become a standard backbone. Further developments include IMP-GCN \cite{liu2021interest}, which applies GNNs to sub-graphs, and LayerGCN \cite{zhou2023layer}, which incorporates residual networks into GNN architectures. FourierKAN-GCF \cite{xu2024fourierkan} optimizes non-linear transformations using Fourier series.
Beyond traditional recommendation, GNNs have been applied in multimodal \cite{xu2024mentor,guo2024lgmrec,zhou2023tale,wei2023multi}, social \cite{fan2019graph,quan2023robust,wang2019neural,li2024recdiff}, and group \cite{xu2024aligngroup,chen2022thinking,wu2023consrec,sankar2020groupim} recommendations to leverage the structural richness of graphs. However, GNN-based methods often rely heavily on supervised signals, limiting their performance on sparse datasets.
\subsection{CL-based Recommendation}
Contrastive Learning (CL) addresses data sparsity in recommendation by leveraging self-supervised signals to construct contrastive views and maximize mutual information in embeddings. SGL \cite{wu2021self} ensures consistency between views using random corruption operators, such as node/edge deletion and random walks. SimGCL \cite{yu2022graph} adds noise during graph convolution, while LightGCL \cite{cailightgcl} employs Singular Value Decomposition (SVD) to generate views. HCCF \cite{xia2022hypergraph} integrates global information via hypergraph neural networks. NCL \cite{lin2022improving} introduces EM clustering, and DCCF \cite{ren2023disentangled} learns disentangled representations using self-supervised signals. BIGCF \cite{zhang2024exploring} explores the individuality and collectivity of user intents with tailored signals.
While effective, existing CL methods incur additional computational costs and introduce irrelevant noise during contrastive view construction. Our NLGCL leverages naturally existing contrastive views within GNNs, reducing computational overhead and ensuring that the noise between contrastive views is semantically related, thereby achieving both improvements in effectiveness and efficiency.
\section{Conclusion}
In this paper, we propose a simple yet effective paradigm, NLGCL, which leverages naturally existing contrastive views between neighbor layers within GNNs. By treating each node and its neighbors in the next layer as positive pairs and other nodes as negatives, NLGCL avoids augmentation-based noise while preserving semantic relevance. This approach reduces irrelevant noise and eliminates the additional time and space costs of traditional contrastive learning methods. We demonstrate the superiority of NLGCL over state-of-the-art baselines in terms of both effectiveness and efficiency through theoretical analysis and empirical evaluation. By eliminating the need to construct and store additional contrastive views, NLGCL breaks the limitations of traditional GCL-based recommendation paradigms and opens new avenues for research. In the future, we plan to extend this paradigm to other domains, such as multimodal, social, and group recommendations.

\begin{acks}
This work was supported by the Hong Kong UGC General Research Fund no. 17203320 and 17209822, and the project grants from the HKU-SCF FinTech Academy.
\end{acks}

\appendix
\section{Appendix}

\subsection{Proof of Theorem~\ref{th:1}}
\label{appendix: proof}
\textbf{\textit{Formal Restatement:}} In GNNs, the effectiveness of naturally existing contrastive views between adjacent layers $l-1$ and $l$ diminishes as the layer index $l$ increases. Specifically, let the information gain $\mathbf{I}(\mathbf{E}^{(l-1)} ; \mathbf{E}^{(l)})$ denote the mutual information between embeddings of adjacent layers. Then, monotonically decreases as $l$ increases.

\noindent\textbf{\textit{Notation:}} \textbf{(Mutual Information)} Define mutual information between adjacent layers as:
\begin{equation}
    \mathbf{I}(\mathbf{E}^{(l-1)} ; \mathbf{E}^{(l)})=\mathbf{H}(\mathbf{E}^{(l-1)})-\mathbf{H}(\mathbf{E}^{(l-1)} | \mathbf{E}^{(l)}),
\end{equation}
where $\mathbf{H}(\cdot)$ denotes the entropy function.

\noindent \textbf{(Spectral Decomposition)} Let $\tilde{\mathcal{A}}=\mathbf{U} \Lambda \mathbf{U}^{\top}$ be the eigendecomposition of $\tilde{\mathcal{A}}$ where $\Lambda=\operatorname{diag}\left(\lambda_1, \ldots, \lambda_n\right)$ and $\lambda_1 \geq \lambda_2 \geq \cdots \geq \lambda_n$

\begin{lemma}\textbf{(Entropy Reduction via Low-Pass Filtering).}
\label{lem:1}
For any layer $l$, the entropy of embeddings satisfies:
\begin{equation*}
\mathbf{H}(\mathbf{E}^{(l)}) \leq \mathbf{H}(\mathbf{E}^{(l-1)})
\end{equation*}
\end{lemma}
\noindent\textbf{\textit{Proof:}} Propagation process $\mathbf{E}^{(l)}=\tilde{\mathcal{A}} \mathbf{E}^{(l-1)}$ acts as a low-pass filter in the spectral domain. Specifically, the frequency response at the $k$-th eigencomponent is attenuated by $\lambda_k^l$. Since $\lambda_k \leq 1$ for all $k$, higher-frequency components (associated with $\lambda_k < 1$) are suppressed exponentially with $l$, reducing the variance of $\mathbf{E}^{(l)}$. From the entropy-power inequality \cite{cover1999elements}:
\begin{equation}
\begin{aligned}
\mathbf{H}(\mathbf{E}^{(l)})&=\frac{1}{2} \log ((2 \pi e)^d|\Sigma^{(l)}|) \\&\leq \frac{1}{2} \log ((2 \pi e)^d|\Sigma^{(l-1)}|)=\mathbf{H}(\mathbf{E}^{(l-1)}),
\end{aligned}
\end{equation}
where $|\Sigma^{(l)}|$ is the determinant of the covariance matrix, and $d$ is the embedding dimension.
\begin{corollary}\textbf{(Exponential Mutual Information Decay).}
\label{coroll:1}
The mutual information between adjacent layers decays as:
\begin{equation*}
\mathbf{I}(\mathbf{E}^{(l-1)} ; \mathbf{E}^{(l)}) \propto \lambda_{\max }^{2 l},
\end{equation*}
where $\lambda_{\max }=\max _k|\lambda_k|<1$.
\end{corollary}
\noindent\textbf{\textit{Proof:}} From Lemma~\ref{lem:1}, $\mathbf{H}(\mathbf{E}^{(l)})$ decreases monotonically. For a Gaussian embedding distribution, mutual information simplifies to:
\begin{equation}
\mathbf{I}(\mathbf{E}^{(l-1)} ; \mathbf{E}^{(l)})=\frac{1}{2} \log (\frac{|\Sigma^{(l-1)}|}{|\Sigma^{(l)}|}).
\end{equation}

Substituting $\Sigma^{(l)}=\tilde{\mathcal{A}}^2 \Sigma^{(l-1)}$ (from $\mathbf{E}^{(l)}=\tilde{\mathcal{A}} \mathbf{E}^{(l-1)}$), we derive:
\begin{equation}
|\Sigma^{(l)}|=|\tilde{\mathcal{A}}^2||\Sigma^{(l-1)}|=\prod_{k=1}^n \lambda_k^2 \cdot|\Sigma^{(l-1)}|.
\end{equation}

Thus:
\begin{equation}
\mathbf{I}(\mathbf{E}^{(l-1)} ; \mathbf{E}^{(l)})=\frac{1}{2} \log (\prod_{k=1}^n \lambda_k^{-2})=-\sum_{k=1}^n \log \lambda_k.
\end{equation}

For dominant eigenvalues $\lambda_{\max}$, this decays as $\mathcal{O}(\lambda_{\max}^{2 l})$. Therefore, contrastive views constructed from lower layers (l = 1) maximize the signal-to-noise ratio (SNR) for contrastive learning, as they preserve higher mutual information. The SNR of contrastive pairs is proportional to $\mathbf{I}(\mathbf{E}^{(l-1)} ; \mathbf{E}^{(l)})$. From Corollary~\ref{coroll:1}, SNR decays exponentially with $l$, making lower layers preferable.

\begin{table}[!h]
\caption{Comparison of time complexity. -H and -E denote heterogeneous and entire scopes, respectively.}
 \vskip -0.1in
\centering
\label{tab: complexity}
\resizebox{\linewidth}{!}{
    \begin{tabular}{c|c|c|c}
    \hline
    ~ & Encoder & BPR Loss & CL Loss \\ \hline
    LightGCN & $O(2|\mathcal{E}|Ld)$ & $O(2dB)$ & - \\ \hline
    SGL & $O(2(1+2 \hat{\rho})|\mathcal{E}|Ld)$ & $O(2dB)$ & $O(2MdB)$ \\ \hline
    NCL & $O(2(|\mathcal{E}|+K)Ld)$ & $O(2dB)$ & $O(4MdB)$ \\ \hline
    SimGCL & $O(6|\mathcal{E}|Ld)$ & $O(2dB)$ & $O(2MdB)$ \\ \hline
    LightGCL & $O(2(|\mathcal{E}|+q|\mathcal{V}|)Ld)$ & $O(2dB)$ & $O(2MdB)$ \\ \hline
    DCCF & $O(2(|\mathcal{E}|+|\mathcal{K}||\mathcal{V}|)Ld)$ & $O(2dB)$ & $O(6MdB)$ \\ \hline
    BIGCF & $O(2(|\mathcal{E}|Ld+|\mathcal{K}||\mathcal{V}|d))$ & $O(2dB)$ & $O(6MdB)$ \\ \hline
    
    NLGCL-H & $O(2|\mathcal{E}|Ld)$ & $O(2dB)$ & $O(2GMdB)$ \\ \hline
    NLGCL-E & $O(2|\mathcal{E}|Ld)$ & $O(2dB)$ & $O(4GMdB)$ \\ \hline
        
    \end{tabular}
    }
    
\end{table}

\subsection{Efficiency Analysis of NLGCL}
\label{appendix: analysis efficiency}
We analyze the complexity of NLGCL and compare it with LightGCN, SGL, NCL, SimGCL, LightGCL, DCCF, and BIGCF. The discussion is within a single batch since the in-batch negative sampling is a widely used trick in CL \cite{chen2020simple}. As Table~\ref{tab: complexity} shows, we divide computational complexity into three components: \textbf{Encoder}, \textbf{BPR Loss}, and \textbf{CL Loss}. LightGCN is the basic graph encoder, the time complexity of the process is $O(2|\mathcal{E}|Ld)$, where $|\mathcal{E}|$ is the number of edges in graph $\mathcal{G}$, $M$ represents the node number in a batch, $L$ is layer number, and $d$ is the dimension of embeddings. $\hat{\rho}$ is the edge keep probability of SGL. $K$ is the number of clusters in NCL. $q$ is the required rank for SVD in LightGCL. $|\mathcal{K}|$ denotes the collective intent number for all user and item nodes. For our NLGCL, $G$ is the number of groups of contrastive views. NLGCL-E requires double samples compared to NLGCL-H in CL loss. In Section~\ref{subsec: effectiveness} and Section~\ref{subsec: efficiency}, we empirically verify that NLGCL-H achieves superior results than NLGCL-E in terms of both effectiveness and efficiency. Since NLGCL-H does not construct additional contrast views through data augmentation, there is no additional cost to the Encoder beyond the backbone LightGCN. For CL Loss, it requires $G$ times the computational cost of other CL-based models due to the multiple groups of natural contrast views. Note that we also empirically prove that (In Section~\ref{subsec: efficiency}): When $G=1$, the performance of NLGCL-H has exceeded the existing baselines.





\newpage


\end{document}